\documentclass[prd,twocolumn,superscriptaddress,nofootinbib,aps,10pt,preprintnumbers]{revtex4-1}

\usepackage[colorlinks=true,linkcolor=black,citecolor=blue,urlcolor=blue, pdfborder={0 0 0}]{hyperref}
\usepackage{mathtools}
\usepackage[utf8]{inputenc}

\def\eq#1{{Eq.~(\ref{#1})}}
\def\eqs#1#2{{Eqs.~(\ref{#1})--(\ref{#2})}}

\renewcommand{\bar}{\overline}
\newcommand{\N}{\mathcal{N}}
\newcommand{\mX}{\mathcal{X}}

\begin{document}

\preprint{CERN-TH-2019-118}
\title{Axion-electron decoupling in nucleophobic axion models}

\newcommand{\affPisa}{{\small \it Dipartimento di Fisica, Universit\`a di Pisa and INFN, Sezione di Pisa, \\ Largo B. Pontecorvo 3, 56127 Pisa, Italy}}
\newcommand{\affINFN}{{\small \it INFN, Laboratori Nazionali di Frascati, C.P.~13, 100044 Frascati, Italy}}
\newcommand{\affBarcelona}{{\small \it Dept.~de F\'{\i}sica Qu\`antica i Astrof\'{\i}sica, Institut de Ci\`encies del Cosmos (ICCUB), \\ Universitat de Barcelona, Mart\'i Franqu\`es 1, E08028 Barcelona, Spain}}
\newcommand{\affCERN}{{\small \it Theoretical Physics Department, CERN, Geneva, Switzerland}}
\newcommand{\affGranSasso}{{\small \it Laboratori Nazionali del Gran Sasso, Via G. Acitelli, 22, I-67100 Assergi (AQ), Italy}}

\author{Fredrik Bj\"orkeroth}
\affiliation{\affINFN}

\author{Luca Di Luzio}
\affiliation{\affPisa}

\author{Federico Mescia}
\affiliation{\affBarcelona}

\author{Enrico Nardi}
\affiliation{\affINFN}

\author{Paolo Panci}
\affiliation{\affPisa}
\affiliation{\affCERN}
\affiliation{\affGranSasso}

\author{Robert Ziegler}
\affiliation{\affCERN}

\begin{abstract}
  The strongest upper bounds on the axion mass come from astrophysical
  observations like the neutrino burst duration of SN1987A, which
  depends on the axion couplings to nucleons, or the white-dwarf cooling
  rates and red-giant evolution, which involve the axion-electron coupling. 
  It has been recently argued that in variants of DFSZ
  models with generation-dependent Peccei-Quinn charges an
  approximate axion-nucleon decoupling can occur, strongly relaxing
  the SN1987A bound. However, as in standard DFSZ models, the axion
  remains in general coupled to electrons, unless an \emph{ad hoc}
  cancellation is engineered. 
  Here we show that axion-electron decoupling can be 
  implemented without extra tunings in DFSZ-like models with three Higgs
  doublets. Remarkably, the numerical value of the quark mass ratio
  $m_u/m_d\sim 1/2$ is crucial to open up this possibility.
\end{abstract}

\maketitle

\section{Introduction}

It has been recently argued \cite{DiLuzio:2017ogq} that in variants of
Dine-Fischler-Srednicki-Zhitnitsky (DFSZ)
\cite{Dine:1981rt,Zhitnitsky:1980tq} models with two Higgs doublets
and generation-dependent Peccei Quinn (PQ) charges, it is possible to
strongly suppress axion couplings to nucleons (axion nucleophobia).
This implies that the upper limit on the DFSZ axion mass from the neutrino
burst duration of the supernova (SN) SN1987A, which is particularly
strong and generally considered ineludible, can in fact be sizeably
relaxed.%
\footnote{%
  For Kim-Shifman-Vainshtein-Zakharov
  (KSVZ)~\cite{Kim:1979if,Shifman:1979if} axions instead, no
  suppression mechanism for the axion couplings to nucleons can be
  enforced, since they are determined in a model-independent way, and
  yield the often quoted limit
  $m_a \lesssim 0.02\,$eV~\cite{Tanabashi:2018oca}. Note, however,
  that recent analyses of the axion emissivity from the SN core hint
  to a weakening of the bound by a factor of a
  few~\cite{Chang:2018rso,Carenza:2019pxu}.
}
The parameter space region that opens up for nucleophobic axions is, however, only 
marginal. This is because in DFSZ models the axion also couples to
electrons, and then limits from anomalous cooling of white dwarfs and red giants, 
which are only moderately less restrictive, apply. 
On the other hand, generation-dependent PQ charges imply that
the axion couplings to the mass eigenstate fermions receive
corrections from inter-generational mixing. In
Ref.~\cite{DiLuzio:2017ogq} this type of effect was invoked to arrange
for a tuned cancellation between two contributions to the
axion-electron coupling: one proportional to the electron PQ charge,
and the other coming from inter-generational mixing between the
leptons.  This allows the construction of models of nucleophobic \emph{and}
electrophobic axions that can evade all the tightest
astrophysical bounds (astrophobic axions).
Although the tuning of the cancellation required to achieve a
significant level of electrophobia is at the level of 10\%,
astrophobic axion models constructed in this way are admittedly not
particularly elegant.  In this short note we put forth a more natural
way to achieve astrophobia, which requires extending the scalar sector
by a third Higgs doublet, but does not involve any \emph{ad hoc}
cancellation between different contributions to the axion-electron
coupling. Remarkably, this mechanism implies a strong correlation
between the couplings whereby the higher the level of suppression
of the axion-nucleon coupling, the more the axion becomes
electrophobic. Intriguingly, the mechanism can be implemented thanks
to the fact that the light quark mass ratio is close to
$m_u/m_d \approx 1/2$.

\section{The conditions for nucleophobia}

Let us first recall the conditions for nucleophobia. We define the 
axion couplings to nucleons via 
\begin{equation}
\label{eq:axionN}
    \frac{C_\N}{2f_a} \partial_\mu a \, \bar{\N} \gamma^\mu \gamma_5 \N ,
\end{equation}
with $\N=p,n$, while the fundamental couplings of the axion
to quarks $C_q$, with $q=u,d,\dots$ are defined from a similar
expression by replacing $\N \to q$ and
$C_\N \to C_q $.  $C_\N$ can be expressed in
terms of $ C_q $ using non-perturbative inputs from nucleon matrix
elements \cite{diCortona:2015ldu}.  To understand the mechanism behind
axion-nucleon decoupling, it is convenient to consider the two linear
combinations $C_p\pm C_n$ and express them in terms of the $C_q$.
This yields
\begin{align} 
\label{eq:CppCn}
    C_p +C_n &= 0.50\left( C_u  + C_d  - 1\right) - 2 \delta_s ,
\\
\label{eq:CpmCn}
    C_p -C_n &= 1.27\left(C_u  - C_d  - f_{ud}\right) , 
\end{align}
where, in the second line, $f_{ud} = f_u - f_d$, with  
$ f_{u,d} = m_{d,u}/(m_d+m_u)$ the model-independent contributions 
induced by the axion coupling to gluons, chosen in such a way that 
the axion does not mix with $\pi^0$. 
In the first line, $1 = f_u+f_d$ is an exact number, while 
$\delta_s = 0.038 \, C_s + 0.012 \, C_c + 0.009 \, C_b  + 0.0035 \, C_t$ 
is a small $\mathcal{O}(5\%)$ correction dominated by the $s$-quark contribution. 
Nucleophobia requires $ C_p \pm C_n \approx 0$,  which is 
possible in variant DFSZ models with two Higgs doublets $H_{1,2}$
and non-universal PQ charge assignment \cite{DiLuzio:2017ogq}. 
To see this, let us focus on the first generation Yukawa terms
\begin{equation}
    \bar q_1 u_1 H_1 + \bar q_1 d_1 \tilde H_2 ,  
\label{eq:1}
\end{equation}
where $\tilde H_2 =i\sigma_2 H_2^*$. 
The axion couplings to the light quark fields  
(neglecting flavour mixing, which is assumed to be small throughout this 
paper%
\footnote{%
  In the presence of flavour mixing, $C_q \to C_q + \Delta C_q$,  
  where $\Delta C_q$ involves quark mass diagonalization matrices.  
  We refer the reader to \cite{DiLuzio:2017ogq} for details.
}) are
\begin{align}
    C_u &= \frac{1}{2N} \left(\mX_{u_1} - \mX_{q_1}\right)=- \frac{\mX_1}{2N} , \\
    C_d &= \frac{1}{2N} \left(\mX_{d_1} - \mX_{q_1}\right)= \frac{\mX_2}{2N} .
\end{align} 
Here $\mX_{u_1}=\mX(u_1)$, etc.~denote the PQ charges of the fermion fields
while $\mX_{1,2}=\mX(H_{1,2})$. The coefficient of the PQ colour anomaly is then
\begin{equation} 
    2N = \sum_{i=1}^3 \left(\mX_{u_{i}} + \mX_{d_{i}} - 2 \mX_{q_{i}} \right) .
\end{equation}
It is also convenient to define the contribution to the color anomaly
from light quarks only:
\begin{equation}
    2N_\ell = \mX_{u_1} + \mX_{d_1} - 2 \mX_{q_1} = \mX_2-\mX_1 .
\end{equation}
The first condition for ensuring approximate nucleophobia then reads
(cf.~\eq{eq:CppCn})
\begin{equation}
\label{eq:CupCd}
    C_u + C_d = \frac{N_\ell}{N} = 1,
\end{equation}
i.e. only models in which the color anomaly is determined solely by
the light $u,d$ quarks (while the contributions from the two heavier
generations cancel or vanish identically) have a chance to be
nucleophobic.%
\footnote{%
  It is worthwhile mentioning that a certain number of models sharing 
  precisely this property were found in a recent study of $U(1)$ 
  flavour symmetry for the quark sector~\cite{Bjorkeroth:2018ipq}.
}
As emphasized in~\cite{DiLuzio:2017ogq}, this implies that nucleophobic 
axion models can be realized only if the PQ charges are generation-dependent.

Assuming that the first condition \eq{eq:CupCd} is satisfied, the
second condition (cf.~\eq{eq:CpmCn}) reads
\begin{equation}
\label{eq:CumCd}
    C_u - C_d = \frac{-\mX_1 - \mX_2}{2N} = \frac{\mX_1 + \mX_2}{\mX_1 - \mX_2} = f_{ud} ,
\end{equation}
that is,
\begin{equation}
\label{eq:X1divX2}
    \frac{\mX_1}{\mX_2} = -\frac{m_d}{m_u}.
\end{equation}
Finally, by imposing the condition which ensures that the physical
axion field is orthogonal to the Goldstone field of hypercharge $U(1)_Y$, i.e.
$\mX_1 v^2_1 + \mX_2 v^2_2 = 0 $, we obtain a relation between the
ratio of the vacuum expectation values (VEVs) $v_{1,2} = \langle H_{1,2}\rangle$ and the ratio 
of the quark masses that must be satisfied in order to ensure
nucleophobia:
\begin{equation}
\label{eq:V2divV1}
    \frac{v^2_2}{v^2_1} = -\frac{\mX_1}{\mX_2} = \frac{m_d}{m_u}.
\end{equation} 
With only two Higgs doublets responsible for breaking the electroweak
symmetry and for providing masses to all the fermions, the lepton
sector is unavoidably charged under the PQ symmetry and, as mentioned
in the introduction, electrophobia can only be enforced by tuning a
cancellation between the contribution to the axion electron coupling
proportional to the  electron PQ charge, and corrections arising from
lepton flavour mixing~\cite{DiLuzio:2017ogq}.  A possible, and more
elegant alternative, is to introduce a third Higgs doublet $H_3$ with
PQ charge $\mX_3$ that couples only to the leptons, and verify if the
condition $\mX_3 \approx 0$ can be consistently implemented.  In this
case the whole lepton sector would be approximately neutral under the
PQ symmetry, and in particular the axion would decouple from the
electrons.  This possibility is explored in the remainder of the paper.

\section{The conditions for electrophobia} 

Let us consider a three-Higgs doublet model (3HDM) wherein $ H_{1,2} $
couple to quarks as above, while $ H_3 $ couples to the leptons.  We
want to study if electrophobia can be implemented consistently with
nucleophobia.  Assuming \eq{eq:CupCd} is satisfied, we have four
additional conditions: orthogonality between the physical axion and
the hypercharge Goldstone, the second condition for nucleophobia
\eq{eq:X1divX2}, and two conditions on the PQ charges that follow from
the requirement that the four $U(1)$ rephasing symmetries of the
kinetic term of the four scalar fields $H_{1,2,3}$ and $\phi$ (the
latter being the Standard Model (SM) singlet, with $\mX_\phi = 1$, 
responsible for PQ breaking) are broken down to 
$U(1)_Y\times U(1)_{PQ}$.  These last two
conditions can be implemented either by coupling the leptonic Higgs
doublet $H_3$ to both hadronic Higgses ($H_{1,2}$), or by coupling one
of the two hadronic Higgses to the other two doublets:
\begin{equation}
\label{eq:H1H2H3}
    H_3^\dagger H_1 \phi^m + H_3^\dagger H_2 \phi^n 
    ~~\mathrm{or}~~  
    H_3^\dagger H_{1,2} \phi^m + H_{2}^\dagger H_{1} \phi^n . 
\end{equation}
For renormalizable operators one has, without loss of generality,
$m=1,2$ and $n=\pm1,\pm2$, where negative values of $n$ mean
Hermitian conjugation $\phi^{n} \equiv (\phi^\dagger)^{|n|}$.  All in
all, for the first case in \eq{eq:H1H2H3} the four conditions read
\begin{align}
\label{eq:nucleoph}
    \frac{\mX_1}{\mX_2} & =-\frac{m_d}{m_u} , \\
\label{eq:3vevs}
    \mX_1 v^2_1 + \mX_2 v_2^2 + \mX_3 v_3^2  & = 0 , \\
\label{eq:m}
    -\mX_3 +\mX_1 + m  & =0 , \\
\label{eq:n}
    -\mX_3 +\mX_2 + n  & = 0 . 
\end{align}
To see if there is a consistent charge assignment that allows us
to decouple the axion from the leptons, let us set $\mX_3 \to 0$.
In this limit, \eq{eq:3vevs} reduces to the previous condition, 
\eq{eq:V2divV1}, while \eqs{eq:m}{eq:n} imply 
$\mX_1 / \mX_2 = m / n$ which, 
together with \eq{eq:nucleoph}, yields
\begin{equation}
    \frac{m_d}{m_u} = - \frac{m}{n} .
\end{equation}
Hence, with the first choice of operators in~\eq{eq:H1H2H3},
electrophobia can be consistently implemented for the following values
of the light quark mass ratio: $m_d/m_u= 2,\,1,\,1/2$.  It is a
fortunate coincidence that the actual value
$m_u/m_d = 0.48(3)$ is perfectly compatible with the first possibility. 
This renders it possible to have electrophobia together with nucleophobia 
by means of a suitable assignment of PQ charges, rather than by tuning of some parameters.
By contrast, if the breaking $U(1)^4 \to U(1)_Y\times U(1)_{PQ}$ is
enforced via the second set of operators in~\eq{eq:H1H2H3},
respectively with $H_1$ or $H_2$ in the first term, electrophobia
would require $m_d/m_u=1$ or $\infty$ in the first case, and
$m_d/m_u=1$ or $0$ in the latter, implying that for both these cases
electrophobia would not be compatible with nucleophobia.

\section{Astrophobic axions in 3HDM models}

In the previous sections we have spelled out which conditions need to
be satisfied to enforce axion-nucleon and axion-electron
decoupling. Clearly, in a realistic scenario we expect that these
conditions are realized only at some level of approximation, so that
$C_{p,n,e}$, rather than vanish, will just be suppressed.  We will now
study more quantitatively the interrelation between nucleophobia and
electrophobia, and the conditions to realize astrophobia with different
levels of accuracy. 

\begin{figure}[htb]
\centering
    \includegraphics[width=\columnwidth]{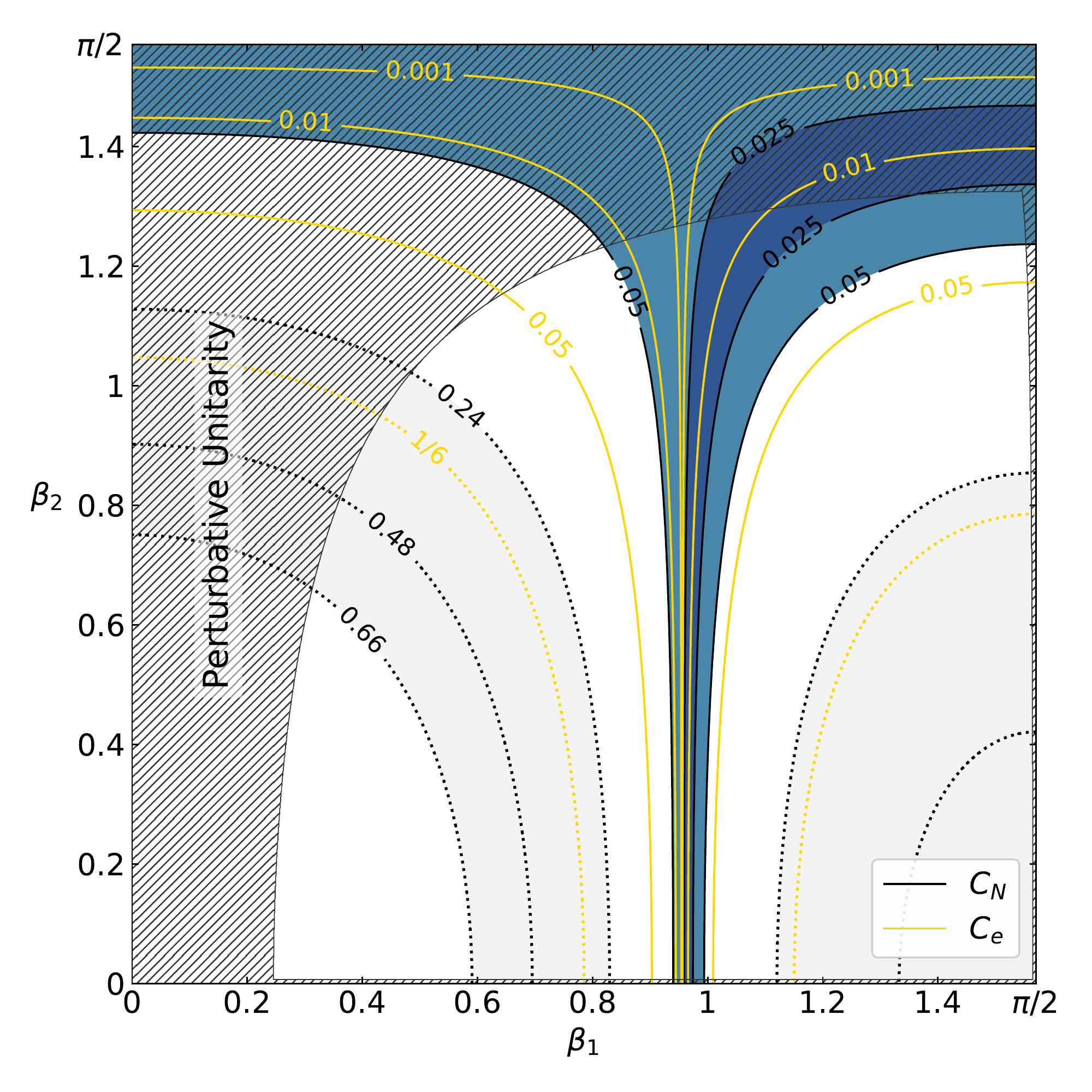}
\caption{%
    Contour lines for $C_N = \sqrt{C_p^2+C^2_n}$ (black) and $C_e$
    (yellow) in the $(\beta_1,\beta_2)$ plane for the astrophobic
    model.  For reference, contour lines corresponding to the values
    of $C_N$ and $C_e$ for the KSVZ and the DFSZ axion models are also
    plotted: $0.48\to C^\mathrm{KSVZ}_N $ (dotted line),
    $[0.24, 0.66]\to
    [C^\mathrm{DFSZ}_{N,\text{min}},\,C^\mathrm{DFSZ}_{N,\text{max}}]$
    (grey region) and $1/6\to C^\mathrm{DFSZ}_{e,(v_1=v_2)}$ (dotted
    yellow line) \cite{Tanabashi:2018oca}.}
\label{fig:bounds}
\end{figure}

According to the results of  the previous section, let us assume that the
scalar potential contains the following terms:
\begin{equation}
\label{eq:H1H2H3exp}
    H_3^\dagger H_1 \phi^2 + H_3^\dagger H_2 \phi^{\dag} . 
\end{equation}
This corresponds to take $m=2$ and $n=-1$ in \eq{eq:H1H2H3} (first
case).  For the quarks we assume a $2+1$ structure with the PQ
charges for the first generation equal to the ones of
the second generation, as in model $(i_{1})$
in~\cite{DiLuzio:2017ogq}.  It is then sufficient to list the Yukawa
operators involving just the second and third generations:
\begin{equation}
\begin{aligned}
    \bar q_2 u_2 H_1, && \bar q_3 u_3 H_2,  &&
    \bar q_2 u_3 H_1, && \bar q_3 u_2 H_{2}, \\
    \bar q_2 d_2 \tilde H_2, && \bar q_3 d_3 \tilde H_1,  &&
    \bar q_2 d_3 \tilde H_{2}, && \bar q_3 d_2 \tilde H_{1} .
\end{aligned}
\end{equation}
We now assume that the leptons couple to a third Higgs doublet with  
the same charges for all generations:
\begin{equation}
    \bar \ell_i e_j \tilde H_3.
\end{equation}
\eqs{eq:m}{eq:n} imply $\mX_1=\mX_3-2$ and $\mX_2=\mX_3+1$.
Neglecting flavour mixings, the diagonal axion couplings to quarks and
leptons read
\begin{equation}
\begin{aligned}
\label{eq:Cu}
    C_{u,c}     &= \frac{2}{3} - \frac{\mX_3}{3}, &\quad 
    C_t         &=-\frac{1}{3} - \frac{\mX_3}{3}, \\
    C_{d,s}     &= \frac{1}{3} + \frac{\mX_3}{3}, & 
    C_b         &= -\frac{2}{3} + \frac{\mX_3}{3} ,\\
    C_{e,\mu,\tau} &= \frac{\mX_ 3}{3} . &&
\end{aligned}
\end{equation}
The nucleophobic condition in \eq{eq:CumCd} then reads 
\begin{equation}
\label{eqmin:chi3}
    \mX_3=\frac{1}{2}-\frac{3}{2}f_{ud} \approx -0.03 , 
\end{equation}
which, as already anticipated, automatically yields a suppressed coupling to electrons:  $C_e\approx 0.01$ 
(for comparison, in DFSZ models with $v_2 = v_1$ one has $C_e = 1/6$).  
The SN bound can be expressed as a constraint on the quantity
$C_N=\sqrt{C_p^2+C^2_n}$ ~\cite{Giannotti:2017hny,Tanabashi:2018oca}
that, according to \eqs{eq:CppCn}{eq:CpmCn}, has a lowest value
$ C_N \approx 0.019$  
which is determined by the correction $\delta_s$ in
\eq{eq:CppCn} (for comparisons in KSVZ axion models
$ C_N \approx 0.48$).%
\footnote{%
  An extra tuning with flavour mixings in~\eq{eq:Cu} can in principle 
  compensate for $\delta_s$ and further reduce the value of $C_N$, 
  see \cite{DiLuzio:2017ogq} for details.
} 
To identify the parameter space regions corresponding
to a sizeable suppression of the  couplings, 
and to show the parametric correlation between nucleophobia and 
electrophobia, it is convenient to parameterize the VEVs as 
\begin{align}
\begin{split}
\label{eq:s1s2}
    v_1 = v c_1 c_2 , \quad
    v_2 = v s_1 c_2 , \quad
    v_3 = v s_2 , 
\end{split}
\end{align}
where $s_i=\sin\beta_i$ and $c_i=\cos\beta_i$, and plot the value of
$C_N$ and $C_e$ as a function of the angles $\beta_{1,2}$ rather than
in terms of the VEVs ratios $\tan \beta_1=v_2/v_1$ and $\tan
\beta_2=\sqrt{v^2_3/(v^2_2+v^2_1)}$.  Although the latter are the
relevant physical parameters, this has the virtue of zooming in on the
regions in which the values of the VEVs are not strongly hierarchical
($\tan\beta_{1,2} \sim O(1)$), and highlighting the correlation between
electrophobia and nucleophobia.  In the
parametrization of \eq{eq:s1s2}, the orthogonality condition in \eq{eq:3vevs}
reads $\mX_3 = (3 c_1^2 - 1) c^2_2$. 
The requirement  that the Yukawa couplings in the 3HDM remain perturbative
restricts the allowed region  in the $(\beta_1, \beta_2)$ plane.
A conservative limit is obtained by imposing the tree-level unitarity bound 
on the $2 \to 2$ fermion scattering, $|\text{Re} \ a_{J=0}| < 1/2$, in the 
3HDM theory involving Yukawas at $\sqrt{s} \gg M_{H_{1,2,3}}$.  
Ignoring running effects, which would make the bound somewhat stronger, and 
taking into account group theory factors 
(see e.g.~\cite{DiLuzio:2016sur,DiLuzio:2017chi}) we get: 
$y^\mathrm{3HDM}_{t,b} < \sqrt{16\pi/3}$ 
(from $Q_L \bar u_R \to Q_L \bar u_R $, with the initial and final states prepared into an $SU(3)_c$ singlet) and 
$y^\mathrm{3HDM}_{\tau} < \sqrt{4\sqrt{2}\pi}$ 
(from $Q_L \bar Q_L \to u_R \bar u_R$, with the initial state prepared into an $SU(2)_L$ singlet).  
The label 3HDM reminds us that these are not the Yukawa couplings of the
SM. The latter are related to the former via:
$y_t = y^\mathrm{3HDM}_t s_1 c_2$, $y_b = y^\mathrm{3HDM}_b c_1 c_2$,
$y_\tau = y^\mathrm{3HDM}_\tau s_2$.  The unitarity bounds on
$y^\mathrm{3HDM}_{t,b,\tau}$ can be now translated into a perturbativity
bound in the $(\beta_1, \beta_2)$ plane, and this results in the
hatched region in Fig.~\ref{fig:bounds}.
Contour lines for different values of $C_{N}$ and $C_e$ are also
plotted in Fig.~\ref{fig:bounds}, and show how electrophobia and
nucleophobia occur in overlapping regions, so that a single choice of
the values of the relevant parameters simultaneously realizes both
properties. It should be also noted that while for small values of
$\beta_2$ the region with suppressed couplings is rather narrow, and
astrophobia requires some tuning of the ratio $v_2/v_1$ to values
sufficiently close to $\sqrt{2}$ ($\beta_1\approx 0.95$), at larger
values $\beta_2\sim \mathcal{O}(1)$ the region opens up and less
tuning is required to simultaneously decouple the axion from nucleons
and electrons.

We recall here for completeness that since nucleophobia requires
generation-dependent PQ charges, axion couplings to quarks are in
general flavour-violating. As discussed in~\cite{DiLuzio:2017ogq},
limits on FCNCs such as $K \to \pi a$ yield stringent
constraints on nucleophobic models which can be complementary to
astrophysics bounds.

A final remark about the axion coupling to photons is in order.
The coupling is defined by the interaction term
\begin{equation}
    \frac{\alpha}{8\pi} \frac{C_\gamma}{f_a} a F_{\mu\nu} \tilde F^{\mu\nu} ,
\end{equation}
where $C_\gamma = E/N - 1.92(4)$, with $E$ denoting the coefficient of
the electromagnetic anomaly.  In astrophobic models there is no
particular reason for which $C_\gamma$ should be suppressed.  In the
present model the contributions of the quarks and leptons are respectively 
$E_Q/N = 8/3 - 2 \mX_3$ and $E_L/N = 2 \mX_3$, so that their sum is
$E/N = 8/3$, a value which is often encountered also in other axion
models~\cite{DiLuzio:2016sbl,DiLuzio:2017pfr}.

\section{Conclusions}

Astrophobic axion models, wherein the axion couplings to nucleons and
electrons can be simultaneously suppressed well below the values
suggested by well-known benchmark models, can be elegantly implemented
in a variant of the DFSZ model in which the PQ charges of the quarks
are generation-dependent, and the scalar sector contains three Higgs
doublets, one of which couples to the leptons and the other two to the
quarks.  While consistent astrophobic axion models were first
constructed in Ref.~\cite{DiLuzio:2017ogq}, in the original scenario
axion decoupling from the electrons was achieved by means of a tuned
cancellation between two different contributions to the axion-electron
coupling, one proportional to the electron PQ charge, and the other
generated by the mixing of the electron with the leptons of the heavier
generations. The virtue of the 3HDM construction presented here is that 
it avoids the need for this cancellation, and enforces
a strong correlation between the suppressions of the axion couplings to
nucleons and electrons, in such a way that nucleophobia and
electrophobia are simultaneously realized in the same region of
parameter space. This renders less contrived 
the possibility that axions might exhibit astrophobic properties.

\section*{Acknowledgments}
\noindent
F.B. and E.N. are supported by the INFN ``Iniziativa
Specifica'' Theoretical Astroparticle Physics (TAsP-LNF).  The work of
L.D.L. is supported by the ERC grant NEO-NAT.  F.M. is supported by
MINECO grant FPA2016-76005-C2-1-P, by Maria de Maetzu program grant
MDM-2014-0367 of ICCUB and 2017 SGR 929.  F.M. acknowledges the INFN
Laboratori Nazionali di Frascati for hospitality and financial
support.  L.D.L. and E.N. acknowledge hospitality from 
Chicago O'Hare International Airport (ORD) and 
the Aspen Center for Physics (ACP) where this work was concluded.  
ACP is supported by National Science Foundation grant PHY-1607611. 
The participation of L.D.L. and E.N. at the ACP was
supported by the Simons Foundation.

%

\end{document}